\documentclass[useAMS,usenatbib]{biom}
\def\bSig\mathbf{\Sigma}

\providecommand{\tabularnewline}{\\}

\usepackage[figuresright]{rotating}
\usepackage{epstopdf}
\usepackage{amsmath}
\usepackage{amssymb}
\usepackage{graphicx}
\usepackage{esint}
\usepackage[authoryear]{natbib}

\title[Generalized Similarity U]{A Generalized Similarity U Test
 for Multivariate Analysis of Sequencing Data}

\author{Changshuai Wei$^{*}$\email{changshuai.wei@unthsc.edu} \\
	   Department of
Biostatistics and Epidemiology, University of North Texas Health Science
Center,\\
Fort Worth, TX 76107, USA
	   \and 
	   Qing Lu$^{*}$\email{qlu@epi.msu.edu}\\
	   Department of Epidemiology and Biostatistics, Michigan State University,
	   East Lansing, MI, 48824, USA
	   }

\begin{document}


\pagerange{\pageref{firstpage}--\pageref{lastpage}} \pubyear{xxxx}

\volume{xx}
\artmonth{xx}
\doi{xxxxxxx/xxxxxxxxxxxxxx}


\label{firstpage}


\begin{abstract}
Sequencing-based studies are emerging as a major tool for genetic
association studies of complex diseases. These studies pose great
challenges to the traditional statistical methods because of the high-dimensionality
of data and the low frequency of genetic variants. Moreover, there
is a great interest in biology and epidemiology to identify genetic
risk factors contributed to multiple disease phenotypes. The multiple
phenotypes can often follow different distributions, which brings an additional challenge to the current statistical framework. In this paper, we propose
a generalized similarity U test, referred to as GSU. GSU is a similarity-based
test that can handle high-dimensional genotypes and phenotypes. We
studied the theoretical properties of GSU, and provided the efficient
p-value calculation for association test as well as the sample size
and power calculation for the study design. Through simulation, we
found that GSU had advantages over existing methods in terms of power
and robustness to phenotype distributions. Finally,
we used GSU to perform a multivariate analysis of sequencing data
in the Dallas Heart Study and identified a joint association of 4
genes with 5 metabolic related phenotypes.
\end{abstract}

%
%


\maketitle

\section{Introduction}
\label{s:intro}

Genome-wide association studies (GWAS) have made
substantial progress in discovering common genetic variants associated
with complex diseases. Despite such success, a large proportion of
heritability of complex diseases remains unexplained. Converging evidence
has suggested that rare variants with minor allele frequency (MAF)
less than 5\% or 1\% hold promise in accounting for a significant
proportion of the missing heritability \citep{Fay2001,Pritchard2001,Kryukov2007,Boyko2008}.
With the advance of next-generation sequencing (NGS) technology, we have now a unique opportunity to investigate the role of a wider scope of
genetic variants, primarily rare variants, in human diseases. Evidence
in early sequencing studies have already shown that rare variants
played an important role in complex diseases \citep{Cohen2004,Ahituv2007,Ji2008,Romeo2009}.
Although promising, the massive data generated from sequencing studies
poses great challenges on the statistical methods. Rare mutations
are recent mutations, and can be only found in a small fraction of
individuals in the entire population. Even with a large effect size,
a rare variant is hard to be detected because of its low MAF. Moreover,
the massive number of rare variants raises the computational burden
and the multiple comparison issue. Therefore, traditional single-locus
analysis has low power to detect rare variants.

Many new statistical methods have been developed for the sequencing
data in the last few years. Different from the traditional single-locus
analysis, most of the new methods perform a joint association test,
namely, testing the joint effect of a set of single nucleotide variants
(SNVs) on a genomic region, a functional unit (e.g., a gene) or a functional
pathway. The advantage of the joint association test over the single-locus
analysis lies in the fact that, by combining multiple SNVs, not only
the association information is aggregated but also the number of tests
is greatly reduced. The joint association tests can be briefly classified
into two categories: burden tests and non-burden tests. Burden tests
first summarize multiple rare variants into a univariate genetic score,
and then test the association of the summary score with the disease
phenotype.\citep{Morgenthaler2007,Li2008,Madsen2009,Lin2011}. Burden
tests often assume the effect of the multiple variants have similar
magnitude and direction. Non-burden tests, on the other hand, can
take into account of the effect heterogeneity within the SNV-set,
by considering the effect of the multiple variants as random effects
or function of genomic position.\citep{Neale2011,Wu2011,Luo2012}.
Among non-burden methods, sequence kernel association test (SKAT)
shares many nice computational and asymptotic properties with the
variance component score test, and has been very popularly used\citep{Wu2011}.

Most of existing joint association tests are parametric-based methods,
which often rely on certain assumptions (e.g., a normal distribution
assumption). When the assumptions are violated, they are subjected to power
loss or an inflated type I error. Moreover, there lacks the development of new methods
for the multivariate analysis of sequencing data, especially when the phenotype follows different
different distributions (e.g., some phenotypes are binary while
the others are continuous). The study of multiple phenotypes is important
in biomedical research. Many studies collect multiple biochemical
measurements related to a disease of interest. These multiple measurements
evaluate different aspects of the disease, and thus better reflect
the underlying biological mechanism of the disease. It also becomes
popular for human genetic studies to collect and study multiple disease
phenotypes. For instance, studies have identified common gene variants
contributed to co-morbidity of substance dependence\citep{Dick2008}.

To achieve these goals, we propose a Generalized Similarity U test,
referred to as GSU. We use two different similarity measurements to
summarize genetic information and phenotypic information, and then
form a test under the weighted U framework. We derive the asymptotic
distribution of the test statistic and implement the new method in
R. The proposed method has several remarkable
features: 1) it is non-parametric and is thus robust to phenotype
distributions; 2) it can handle multiple different phenotypes (e.g.,
a combination of binary and continuous phenotypes); 3)it has a nice
statistical property and performs well under small sample sizes. Simulation
studies are conducted to compare the type 1 error and the power of
our method with those of several commonly used methods. Finally, we
applied the new method to the Dallas Heart Study to test the association
of 4 candidate genes with 5 metabolic-related phenotypes.

\section{Generalized Similarity U}

\subsection{Weighted U Statistic}

Suppose that $n$ subjects are sequenced in a study, where we are
interested in testing the association of L phenotypes ($y_{i,l}$, $1\leq i\leq n$,
$1\leq l\leq L$) with M genetic variants ($g_{i,m}$, $1\leq i\leq n$,
$1\leq m\leq M$). For each subject $i$, we observe a vector of phenotypes
$y_{i}$ ( $y_{i}=(y_{i,1},y_{i,2},\cdots,y_{i,L})$ ) and a vector
of genotypes $g_{i}$ ( $g_{i}=(g_{i,1},g_{i,2},\cdots,g_{i,M})$).
In the special case when $L=1$ (or $M=1$), it is simplified to a
univariate analysis (or a single-locus analysis). When $L>1$, it
extends to a multivariate analysis. In GSU, we allow multiple
phenotypes to be of different types (e.g., continuous or categorical),
and do not assume any distribution of phenotypes. The number of genetic
variants $M$ and the number of phenotypes $L$ can be larger than
the sample size. For example, the genetic data can be sequencing data (high
demensional genotypes) and the phenotype data can be imaging data (high demensional phenotypes).\\

Given the phenotypes and the genotypes for the subjects $i$ and $j$
,we first define their phenotype similarity $S_{i,j}$ by,

\[
S_{i,j}=h(y_{i},y_{j}),
\]
and define their genetic similarity $K_{i,j}$ by,

\[
K_{i,j}=f(g_{i},g_{j}).
\]
The similarity measurements $h(\cdotp,\cdotp)$ and $f(\cdotp,\cdotp)$
defined above can be of a general form as long as they satisfy the
finite second moment condition (i.e.,$E(h^{2}(Y_{1},Y_{2}))<\infty$
and $E(f^{2}(G_{1},G_{2}))<\infty$ ). We further center the phenotype
similarity by,

\begin{eqnarray}
\tilde{S}_{i,j} & = & \tilde{h}(y_{i},y_{j})\nonumber \\
 & = & h(y_{i},y_{j})-E(h(y_{i},Y_{j})) \nonumber \\
 & & -E(h(Y_{i},y_{j}))+E(h(Y_{i},Y_{j})),\label{eq:Centering}
\end{eqnarray}
and center the genetic similarity, $\tilde{K}_{i,j}=\tilde{f}(g_{i},g_{j})$,
in a similar manner. The generalized similarity U (GSU) is then defined
as the summation of the centered phenotype similarities weighted by
the centered genetic similarities,
\begin{equation}
U=\frac{1}{n(n-1)}\sum_{i\neq j}\tilde{K}_{i,j}\tilde{S}_{i,j},\label{eq:GSU}
\end{equation}
where the $\tilde{K}_{i,j}$ is considered as the weight function
and the $\tilde{S}_{i,j}$ is considered as the U kernel. In our definition
of GSU, the role of genetic similarity and phenotype similarity are
interchangeable. In other words, we can also treat $\tilde{S}_{i,j}$
as the weight function and $\tilde{K}_{i,j}$ as the U kernel.

\subsection{Similarity Measurement}

The choices for the phenotype similarity $h(\cdotp,\cdotp)$ and the
genetic similarity $f(\cdotp,\cdotp)$ are flexible. According to
different types of genetic variants and the purpose of the analysis,
we can choose different types of phenotype similarities and genetic
similarities.

For the categorical SNVs data, we can use either the IBS function
or the weighted IBS function to measure the genetic similarity\citep{Lynch1999}.
Assuming the genetic variants ($g_{i,m}$, $1\leq i\leq n$, $1\leq m\leq M$)
are coded as 0, 1 and 2, respectively for AA, Aa and aa, we can define
the IBS-based genetic similarity between subjects $i$ and $j$ as,
\[
K_{i,j}^{IBS}=\frac{1}{2M}\sum_{m=1}^{M}2-|g_{i,m}-g_{j,m}|.
\]
Alternatively, we can use a weighted-IBS (wIBS) genetic similarity
to emphasize the effects of rare variants,
\[
K_{i,j}^{wIBS}=\sum_{m=1}^{M}\frac{w_{m}(2-|g_{i,m}-g_{j,m}|)}{\Upsilon},
\]
where $w_{m}$ represents the weight for the $m$-th SNV in the SNV-set,
and $\Upsilon$ is a scaling constant, defined as $\Upsilon=2\sum_{m=1}^{M}w_{m}$.
$w_{m}$ is usually defined as a function of minor allele frequency (MAF, denoted as $\gamma_{m}$). For
example, we can define the weight $w_{m}$ as $w_{m}=1/\sqrt{\gamma_{m}(1-\gamma_{m})}$
. When the genetic data are count data or continuous data, we can
use other forms of $f(\cdotp,\cdotp)$ to measure genetic similarity
(e.g., euclidian distance based similarity), which, we will leave
for further investigations.

For phenotype similarity, we define a unified measurement for both
categorical and continuous phenotypes based on a normal quantile.
For each phenotype, $y_{l}$ ($1\leq l\leq L$), we define the corresponding
normal quantile by,
\[
q_{i,l}=\Phi^{-1}((rank(y_{i,l})-0.5)/n),
\]
where $rank(\cdotp)$ corresponds to the rank of the phenotype value
$y_{i,l}$. When there are ties, we assign the averaged rank. $\Phi^{-1}(\cdotp)$
is the inverse cumulative density function for a standard normal distribution.
We can then calculate the Euclidian Distance (ED) based phenotype
similarity between subjects $i$ and $j$ by,
\[
S_{i,j}^{ED}=exp(-\sum_{l=1}^{L}\omega_{l}(q_{i,l}-q_{j,l})^{2}),
\]
where $\omega_{l}$ represents the weight for the $l$-th phenotypes given based on prior knowledge. If there is no prior
knowledge, we can use an equal weight, $\omega_{l}=1/L$. The ED-based
phenotype similarity can be easily modified to take the correlation
among the phenotypes into account,
\[
S_{i,j}^{ED}=exp\left(-\frac{1}{L}(q_{i}-q_{j})^{T}\Gamma(q_{i}-q_{j})\right),
\]
where $q_{i}=(q_{i1},\cdots,q_{iL})^{T}$. $\Gamma$ can be chosen
to reflect the correlations among the phenotypes. For example, we
can define $\Gamma$ as,
\[
\Gamma=(\frac{1}{n}\sum_{i=1}^{n}q_{i}q_{i}^{T})^{-1}.
\]
Other than the ED based phenotype similarity, we can define the phenotype
similarity using cross product\citep{Tzeng2009}. The types of phenotype
similarity or genetic similarity can be chosen based on different
purposes, which may influence the power of the method. For simplicity,
we used an ED-based phenotype similarity in this paper.

\subsection{Hypothesis testing}

Based on the definition of the centered similarity, we can show that
$E(\tilde{f}(G_{i},G_{j}))=0$ and $E(\tilde{h}(Y_{i},Y_{j}))=0$
(Appendix A). Under the null hypothesis,
when the genetic variants are not associated with multiple phenotypes,
we have $E(U)=0$ (Appendix A). Under
the alternative hypothesis, when the genetic variants are associated
with multiple phenotypes, we expect that the phenotype similarity
is concordant with the genetic similarity. In other words, the positive
phenotype similarities are weighted heavier and the negative phenotype
similarities are weighted lighter, leading to a positive value of
U statistic. A statistical test can be formed to test the association
and its p-value can be calculated by $P(U>U_{obs})$, where $U_{obs}$
is the observed value of $U$. Tests based on permutation or bootstrap
can be implemented to obtain the statistical significance.
Nevertheless, both methods are computationally expensive for high-dimensional
data. Therefore, we derive the asymptotic distribution of GSU to assess
the statistical significance of the association test. \\

By considering the genetic similarity as the weight function and the
phenotype similarity as the U kernel, GSU is a weighted U statistic\citep{Gregory1977,Shapiro1979,ONeil1993,Shieh1997,Lindsay2008}.
More specifically, because its kernel satisfied $Var(E(\tilde{h}(Y_{1},Y_{2})|Y_{2}))=0$
(Appendix A), GSU is a degenerated weighted
U statistic. To derive the limiting distribution of GSU, we can decompose
the centered phenotype similarity by,
\[
\tilde{h}(y_{1},y_{2})=\sum_{s=1}^{\infty}\lambda_{s}\phi_{s}(y_{1})\phi_{s}(y_{2}),
\]
where $\{\lambda_{s}\}$ and $\{\phi_{s}(\cdotp)\}$ are eigenvalues
and eigenfunctions of the U kennel $\tilde{h}(\cdotp,\cdotp)$, and
all the eigenfunctions are orthogonal,
\[
\int\phi_{s}(y_{1})\phi_{s'}(y_{1})dF(y_{1})=\begin{cases}
0, & \text{if \ensuremath{s\neq s'}}\\
1, & \text{if \ensuremath{s=s'.}}
\end{cases}
\]
Similarly, we can decompose the centered genetic similarity by,
\[
\tilde{f}(G_{i},G_{j})=\sum_{t=1}^{\infty}\eta_{t}\varphi_{t}(g_{1})\varphi_{t}(g_{2}).
\]
We can then rewrite the GSU as,

\begin{eqnarray*}
U & = & \frac{1}{n-1}\sum_{t=1}^{\infty}\sum_{s=1}^{\infty}sign(\eta_{t}\lambda_{s})\left(\frac{1}{\sqrt{n}}\sum_{i=1}^{n}\eta_{t}^{\star}(G_{i})\phi_{s}^{\star}(Y_{i})\right)^{2}\\
 &  & -\frac{1}{n-1}\sum_{t=1}^{\infty}\sum_{s=1}^{\infty}sign(\eta_{t}\lambda_{s})\frac{1}{n}\sum_{i=1}^{n}\left(\eta_{t}^{\star}(G_{i})\phi_{s}^{\star}(Y_{i})\right)^{2},
\end{eqnarray*}
where $\varphi_{t}^{\star}(G_{i})=|\eta_{t}|^{0.5}\varphi_{t}(G_{i})$
and $\phi_{s}^{\star}(Y_{i})=|\lambda_{s}|^{0.5}\phi_{s}(Y_{i})$ (Appendix B).

Using the form above, we can show that the limiting distribution of
GSU is a weighted sum of independent chi-square random variables.
This is the result of theorem 1 below, which is proved in Appendix B.

\begin{theorem}
Suppose $E(h^{2}(Y_{1},Y_{2}))<\infty$, $E(f^{2}(G_{1},G_{2}))<\infty$,
and $Y\perp G$. Let $\tilde{h}(Y_{1},Y_{2})$ and $\tilde{f}(G_{1},G_{2})$
be the centered similarities as defined in (\ref{eq:Centering}). Define
U as $U=\frac{1}{n(n-1)}\sum_{i\neq j}\tilde{f}(G_{i},G_{j})\tilde{h}(Y_{i},Y_{j})$.
Then, $nU\xrightarrow{D}\sum_{t=1}^{\infty}\eta_{t}\sum_{s=1}^{\infty}\lambda_{s}(\chi_{st}^{2}-1)$,
where $\{\chi_{st}^{2}\}$ are independent chi-square random variables
with 1 degree of freedom.
\end{theorem}

\subsection{The power and sample size calculation}

In this subsection, we derive the asymptotic distribution of GSU under the
alternative hypothesis, and provide power and sample size calculations
for sequencing association studies.\\

Assume under the alternative hypothesis that $E(\tilde{f}(G_{1},G_{2})\tilde{h}(Y_{1},Y_{2}))=\mu>0$ and
$Var(\tilde{f}(G_{1},G_{2})\tilde{h}(Y_{1},Y_{2})|(G_{2},Y_{2}))=\zeta_{1}>0$
, where $\mu$ measures the strength
of the association. It is easy to show that GSU is an unbiased estimator
of $\mu$,
\[
E(U)=\frac{1}{n(n-1)}\sum_{i\neq j}E(\tilde{f}(G_{i},G_{j})\tilde{h}(Y_{i},Y_{j}))=\mu.
\]
Using the Hoeffding projection, we can show that GSU asymptotically
follows a normal distribution, with mean $\mu$ and variance $4\zeta_{1}/n$.
This is the result of Theorem 2, which is proved in Appendix C.

\begin{theorem}
Let $\tilde{h}(Y_{1},Y_{2})$ and $\tilde{f}(G_{1},G_{2})$
be the centered similarities as defined in (\ref{eq:Centering}).
Suppose Y is associated with G, and the following conditions are
satisfied: $E(\tilde{f}(G_{1},G_{2})\tilde{h}(Y_{1},Y_{2}))=\mu>0$,
$Var(\tilde{f}(G_{1},G_{2})\tilde{h}(Y_{1},Y_{2}))=\zeta_{0}<\infty$,
and $Var(\tilde{f}(G_{1},G_{2})\tilde{h}(Y_{1},Y_{2})|(G_{2},Y_{2}))=\zeta_{1}>0$.
Define $U$ as $U=\frac{1}{n(n-1)}\sum_{i\neq j}\tilde{f}(G_{i},G_{j})\tilde{h}(Y_{i},Y_{j})$.
Then, $\sqrt{n}(U-\mu)\xrightarrow{D}N(0,4\zeta_{1})$.
\end{theorem}

The power of GSU at the significance level $\alpha$ can be calculated
by,
\begin{align*}
P\{nU & >q_{1-\alpha}\}\\
= & P\left\{ \frac{\sqrt{n}(U-\mu)}{2\sqrt{\zeta_{1}}}>\frac{q_{1-\alpha}-n\mu}{2\sqrt{n\zeta_{1}}}\right\} \\
= & \Phi(\frac{n\mu-q_{1-\alpha}}{2\sqrt{n\zeta_{1}}}),
\end{align*}
where $q_{1-\alpha}$ is the $1-\alpha$ quantile for $\sum_{t=1}^{\infty}\eta_{t}\sum_{s=1}^{\infty}\lambda_{s}(\chi_{st}^{2}-1)$
and $\Phi(\cdotp)$ is the CDF of a standard normal distribution.
The sample size required to achieve power $\beta$ can be calculated
by solving $\Phi(\frac{n\mu-q_{1-\alpha}}{2\sqrt{n\zeta_{1}}})\geq\beta$.
By denoting $Z_{\beta}$ as the $\beta$ quantile for a standard normal
distribution, the required sample size is given by,
\[
n=\min_{n\in N}\left\{ n:\: n\geq\frac{\left(Z_{\beta}\sqrt{\zeta_{1}}+\sqrt{Z_{\beta}^{2}\zeta_{1}+\mu q_{1-\alpha}}\right)^{2}}{\mu^{2}}\right\} .
\]

\subsection{Computation and implementation}

Let $S=\{S_{i,j}\}_{n\times n}$ and $K=\{K_{i,j}\}_{n\times n}$
be the matrix form of the phenotype similarity and genetic similarity,
the centered similarity matrices $\tilde{S}$ and $\tilde{K}$ can
be obtained by,

\[
\tilde{S}=(I-J)S(I-J),
\]
\[
\tilde{K}=(I-J)K(I-J),
\]
where $I$ is an n-by-n identity matrix, and $J$ is an n-by-n matrix
with all elements being $1/n$ (Appendix D).
Then GSU can be expressed as,
\[
U=\frac{1}{n(n-1)}\sum_{i\neq j}\tilde{K}_{i,j}\tilde{S}_{i,j}.
\]
In this form, $U$ can be viewed as a sum of the element-wise product
of the two matrices, $\tilde{K}_{0}$ and $\tilde{S}_{0}$, which
are obtained by assigning 0 to the diagonal elements of matrices $\tilde{K}$
and $\tilde{S}$. We then use matrix eigen-decomposition to approximate
the eigen-values in function decomposition. Let $\{\tilde{\lambda}_{s}\}$
and $\{\tilde{\eta}_{t}\}$ respectively be the eigen-values for matrices
$\tilde{K}_{0}$ and $\tilde{S}_{0}$ (Appendix E),
the limiting distribution of $U$ is given by,
\[
nU\sim\sum_{t=1}^{n}\frac{\tilde{\eta}_{t}}{n}\sum_{s=1}^{n}\frac{\tilde{\lambda}_{s}}{n}(\chi_{st}^{2}-1),
\]
where $\{\chi_{st}^{2}\}$ are independent chi-square random variables
with 1 degree of freedom. The p-value can be calculated by using the
Davies' method \citep{Davies1980}, the Liu's method\citep{Liu2009a}
or the Kuonen's method\citep{Kuonen1999}.

\section{Simulation study}

\subsection{Simulation method}

To mimic real genetic structure, we used genetic data from the 1000
Genome Project\citep{Abecasis2010}. Based on the genetic data, we
then simulated phenotype values. In particular, we used a 1Mb region
of the genome (Chromosome 17: 7344328-8344327)
from the 1000 Genome Project. For each simulation replicate, we randomly
chose a 30kb segment from the 1Mb region and formed a SNV-set for
the analysis. From the SNV-set, we
set a proportion of the SNVs as causal. A number of individuals were
randomly chosen from the total 1092 individuals as the simulation sample to study the performance
of the methods. To investigate the robustness
against different phenotype distributions, we simulated three types
of phenotypes: a binary-distributed phenotype, a Gaussian-distributed
phenotype and a Cauchy-distributed phenotype. The binary-distributed
phenotype was simulated by using a logistic regression model,
\[
logit(P(Y_{i}=1))=\mu+G_{i}^{T}\beta,
\]
where $Y_{i}$ and $G_{i}$ were the phenotype value and the genotype
vector (coded as 0, 1, and 2) for the $i$-th individual, respectively.
$\beta$ were the effects of the SNVs, which were sampled from a uniform
distribution with a mean of $\mu_{\beta}$ and a variance of $\sigma_{\beta}^{2}$.
The Gaussian-distributed phenotype was simulated by using the
linear regression model,
\[
Y_{i}=\mu+G_{i}^{T}\beta+\varepsilon_{i},\:\varepsilon_{i}\sim N(0,\sigma^{2}).
\]
The Cauchy-distributed phenotype mimicked a situation in which
the continuous phenotype had more extreme values (i.e. heavy-tailed),
and was simulated by using the following model,
\[
Y_{i}\sim cauchy(a_{i},b),\: a_{i}=\mu+G_{i}^{T}\beta,
\]
where $a_{i}$ and $b$ are the location parameter and the scale parameter
of the Cauchy distribution, respectively.

Two sets of simulation were performed. In simulation I, we considered
a single phenotype, while in simulation II, we considered multiple-phenotypes.

\subsubsection{\textit{Simulation I Setting}}

Under the null, the models were simulated by setting $\mu_{\beta}=0$
and $\sigma_{\beta}^{2}=0$, and by varying the sample
size from 50 to 500 (i.e. 50, 100, 200 and 500). Under the alternative,
two sets of disease models were simulated:
\begin{enumerate}
\item Set $\mu_{\beta}=0$ and $\sigma_{\beta}^{2}>0$ so that half of the
causal SNVs were deleterious and the other half were protective.
\item Set $\mu_{\beta}>0$ and $\sigma_{\beta}^{2}>0$ so that majority
of the causal SNVs were deleterious.
\end{enumerate}

The details of the simulation
setting can be found in Table S1 of Supplementary Materials.

\subsubsection{\textit{Simulation II Setting}}
Two sets of models were simulated:
\begin{enumerate}
\item Assume the multiple phenotypes follow the same distribution. In particular,
we simulated 3 binary-distributed phenotypes (BBB), 3 Gaussian-distributed
phenotypes (GGG), and 3 Cauchy-distributed phenotypes (CCC).
\item Assume the multiple phenotypes follow different distributions.
In particular, we simulated 3 phenotypes with 2 binary-distributed
phenotypes and 1 Gaussian-distributed phenotype (BBG), 3 phenotypes
with 1 binary-distributed phenotypes and 2 Gaussian-distributed phenotypes
(BGG), and 3 phenotypes with 1 binary-distributed phenotype, 1 Gaussian
distributed phenotype and 1 Cauchy distributed phenotype (BGC).
\end{enumerate}
Similar as the simulation I, for the null model, we set $\mu_{\beta}=0$
and $\sigma_{\beta}^{2}=0$ . For the alternative models, we set $\mu_{\beta}=0$
and $\sigma_{\beta}^{2}>0$ , and allowed the multiple phenotypes
to be influenced by different sets of causal SNVs. The details of
the simulation setting were described in Table S2 of Supplementary Materials.

\subsection{Simulation result}

We evaluated the performance of GSU by comparing it with three existing
methods, SKAT\citep{Wu2011}, AdjSKAT and SKATO\citep{Lee2012}. For
each simulation, we created 1000 simulation replicates to evaluate
type 1 error and power. For GSU, we used the wIBS-based genetic similarity
and ED-based phenotype similarity to construct the U statistic. To
be consistent, we used the same weighted IBS to construct the kernel
for SKAT. Because neither AdjSKAT nor SKATO had an option for a weighted IBS
kernel, we used the default kernel, the weighted linear kernel, when
applying these two methods. SKAT, AdjSKAT and SKATO are designed for
univariate analysis. To consider multiple phenotypes, we chose the
most significant p-value from the univariate analysis, and then adjust
for multiple tests by using the Bonferroni correction.

\subsubsection{\textit{Result for Simulation I}}

The type I error rates of the 4 methods are summarized in Table \ref{Tab:1}.
GSU had a well-controlled type I error, regardless of phenotype distributions
and sample sizes. Neverthless, SKAT, AdjSKAT and SKATO had inflated
type I error rates (ranging from 0.101 to 0.19) for the Cauchy-distributed
phenotype. When the sample size was small (e.g., 50 or 100), SKAT and
SKATO also had conservative type I error (e.g., 0.001) for the binary-distributed
phenotype. 

\begin{table}[htbp]
\centering 
\caption{Type I error comparison for the univariate analysis}

\label{Tab:1}

\begin{tabular}{ccccc}
\Hline
Sample size  & Method  & \multicolumn{3}{c}{Distribution}\tabularnewline
\hline
 &  & Binary  & Gaussian  & Cauchy\tabularnewline
\hline
\hline
50  & SKAT  & 0.001  & 0.030  & 0.122\tabularnewline
 & SKATO  & 0.021  & 0.041  & 0.101\tabularnewline
 & AdjSKAT  & 0.054  & 0.028  & 0.123\tabularnewline
 & GSU  & 0.058  & 0.046  & 0.051\tabularnewline
\hline
100  & SKAT  & 0.014  & 0.028  & 0.149\tabularnewline
 & SKATO  & 0.035  & 0.038  & 0.120\tabularnewline
 & AdjSKAT  & 0.063  & 0.027  & 0.139\tabularnewline
 & GSU  & 0.046  & 0.050  & 0.053\tabularnewline
\hline
200  & SKAT  & 0.023  & 0.040  & 0.155\tabularnewline
 & SKATO  & 0.024  & 0.042  & 0.140\tabularnewline
 & AdjSKAT  & 0.043  & 0.042  & 0.156\tabularnewline
 & GSU  & 0.057  & 0.048  & 0.045\tabularnewline
\hline
500  & SKAT  & 0.039  & 0.055  & 0.190\tabularnewline
 & SKATO  & 0.052  & 0.056  & 0.158\tabularnewline
 & AdjSKAT  & 0.049  & 0.053  & 0.180\tabularnewline
 & GSU  & 0.038  & 0.046  & 0.043\tabularnewline
\hline
\end{tabular}
\vspace*{12pt}
\end{table}

\begin{figure*}[htbp]

\centerline{\includegraphics[scale=0.5]{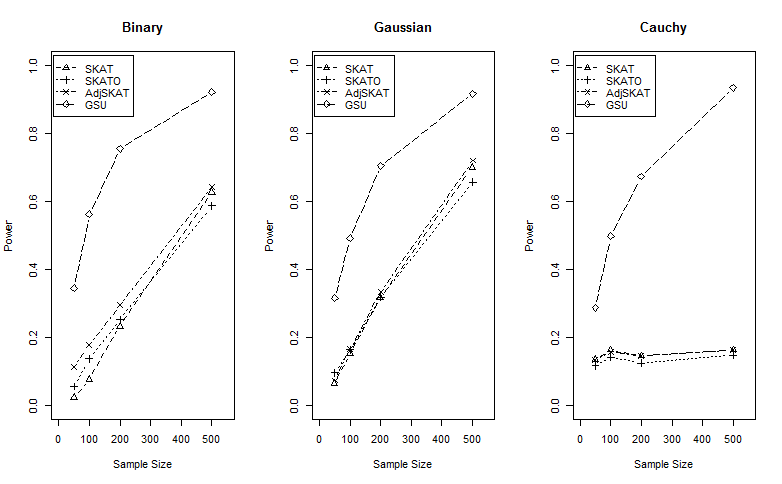}}
\caption{Power comparison for univariate analysis when a majority of causal
SNVs are deleterious}

\label{Fig:1} 

\end{figure*}

\begin{figure*}[htbp]
\centerline{\includegraphics[scale=0.5]{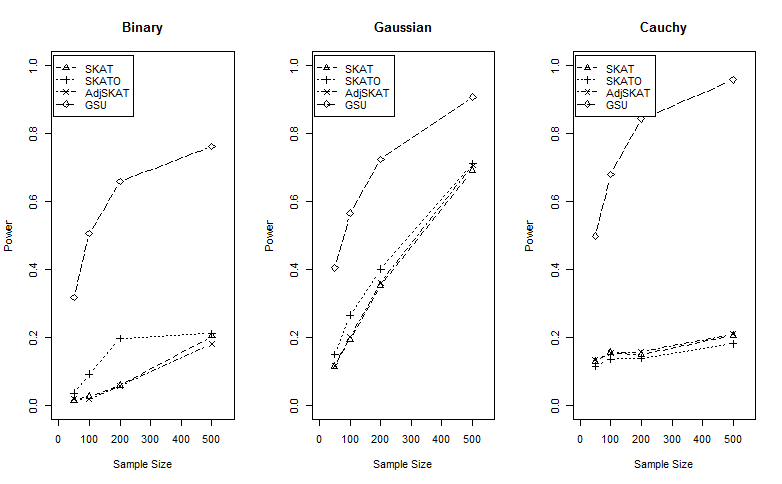}}

\caption{Power comparison for univariate analysis when half of causal SNVs
are protective and the other half of causal SNVs are deleterious}

\label{Fig:2}
\end{figure*}

The power comparison is summarized in Figures \ref{Fig:1} and \ref{Fig:2}.
For the disease model where half of the causal SNVs were deleterious
(Figure \ref{Fig:1}), GSU had higher power than other methods for
various sample sizes ranging from 50 to 500. For instance, in the
simulation with binary phenotype and a sample size of 50, GSU (power=0.346)
attained much higher power than AdjSKAT (power=0.114), SKATO (power=0.057),
and SKAT (power=0.024). Similarly, for the Gaussian-distributed phenotype,
GSU had the highest power among the four methods, and the three SKAT-based
methods had similar performance. For the Cauchy-distributed phenotype,
where the distribution assumption was violated, the power of the SKAT-based
methods remained similar as the sample sizes increased, while GSU
had increased power as the sample size increased. For the second disease
model in which a majority of the SNVs were deleterious (Figure \ref{Fig:2}),
we observed the same conclusion, i.e., GSU had highest power regardless
of sample sizes and phenotype distributions. 

Intuitively, as a semi-parametric model, SKAT should have comparable power with GSU. The reason for the advantage of GSU lies in the form of test statistics. The score test of SKAT utilizes both diagonal terms and off-diagonal terms of the similarity matrix. When the null model gives the same predicted values (i.e., no covariates), the diagonal terms essentially provide no additional information with regard to association, while adding more variation (i.e., noise) to the score statistic. Meanwhile, GSU utilizes only the off-diagonal terms. When the sample size is small, the influence of diagonal terms will be significant, which is why we observed the higher power of GSU over SKAT. When the sample size becomes larger, the off-diagonal terms will dominate, and the difference of power of two methods becomes smaller.

\subsubsection{\textit{Result for Simulation II}}

The type I error rates for the multivariate analysis are summarized
in Table \ref{Tab:2}. Similar to the results of the univariate analysis,
GSU can correctly control type I error at the level of 0.05 (Table
\ref{Tab:2}), while the other three methods had inflated type I error
when the distribution assumption was violated (e.g., CCC and BGC).
When the distribution assumption was satisfied, AdjSKAT correctly
controlled the type 1 error. SKAT and SKATO, however, had conservative
type I error rates for small sample sizes (50 and 100), especially
when the multiple phenotypes comprised of binary-distributed phenotypes
(e.g., BBB, BBG and BGG). 

\begin{table*}[htbp]
\centering
 
\begin{minipage}{130mm}

\caption{Type I error comparison for multivariate analysis}
\label{Tab:2}

\begin{tabular}{cccccccc}
\Hline
Sample size  & Method  & \multicolumn{3}{c}{Same Distr. \footnote{B, G, and C represent Binary-distributed, Gaussian-distributed,
and Cauchy-distributed phenotypes, respectively.} } & \multicolumn{3}{c}{Diff Distr.}\tabularnewline
\hline
 &  & BBB  & GGG  & CCC  & BBG  & BGG  & BGC\tabularnewline
\hline
\hline
50  & SKAT  & 0.002  & 0.028  & 0.232  & 0.008  & 0.021  & 0.097\tabularnewline
 & SKATO  & 0.019  & 0.027  & 0.207  & 0.016  & 0.024  & 0.085\tabularnewline
 & AdjSKAT  & 0.046  & 0.028  & 0.237  & 0.049  & 0.040  & 0.113\tabularnewline
 & GSU  & 0.057  & 0.047  & 0.045  & 0.034  & 0.043  & 0.056\tabularnewline
\hline
100  & SKAT  & 0.001  & 0.023  & 0.295  & 0.011  & 0.013  & 0.122\tabularnewline
 & SKATO  & 0.025  & 0.039  & 0.260  & 0.025  & 0.036  & 0.117\tabularnewline
 & AdjSKAT  & 0.055  & 0.025  & 0.290  & 0.059  & 0.042  & 0.135\tabularnewline
 & GSU  & 0.048  & 0.047  & 0.044  & 0.051  & 0.059  & 0.058\tabularnewline
\hline
200  & SKAT  & 0.020  & 0.048  & 0.325  & 0.033  & 0.020  & 0.134\tabularnewline
 & SKATO  & 0.030  & 0.058  & 0.288  & 0.035  & 0.036  & 0.118\tabularnewline
 & AdjSKAT  & 0.059  & 0.046  & 0.319  & 0.064  & 0.038  & 0.136\tabularnewline
 & GSU  & 0.050  & 0.047  & 0.049  & 0.058  & 0.056  & 0.043\tabularnewline
\hline
500  & SKAT  & 0.044  & 0.035  & 0.364  & 0.035  & 0.04  & 0.166\tabularnewline
 & SKATO  & 0.054  & 0.036  & 0.315  & 0.038  & 0.047  & 0.156\tabularnewline
 & AdjSKAT  & 0.06  & 0.038  & 0.342  & 0.041  & 0.049  & 0.172\tabularnewline
 & GSU  & 0.045  & 0.054  & 0.044  & 0.046  & 0.039  & 0.062\tabularnewline
\hline
\end{tabular}
 
\end{minipage}
\vspace*{12pt}
\end{table*}

\begin{figure*}[htbp]

\centerline{\includegraphics[scale=0.5]{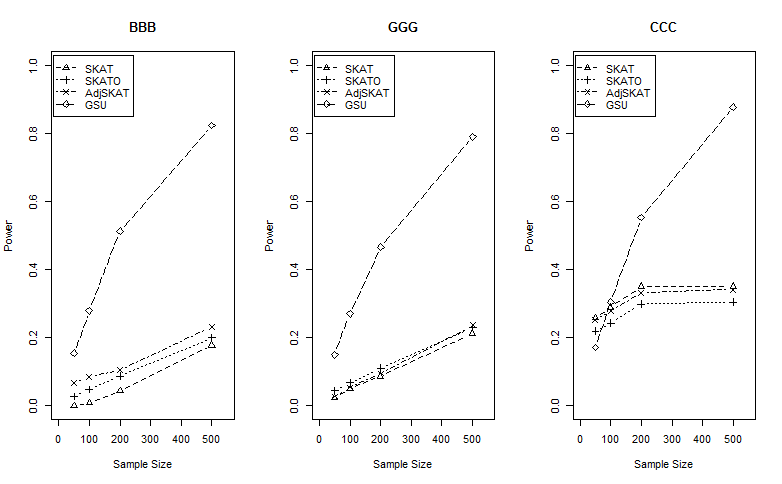}}
\caption{Power comparison for the multivariate analysis when the multiple phenotypes
follow the same type of distribution}

\label{Fig:3} 
\end{figure*}

\begin{figure*}[htbp]
\centerline{\includegraphics[scale=0.5]{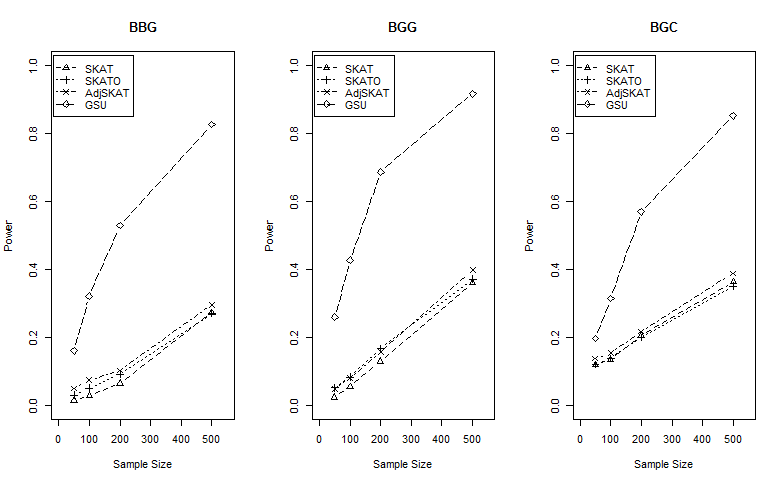}}
\caption{Power comparison for the multivariate analysis when the multiple phenotypes
follow the different types of distributions}

\label{Fig:4} 
\end{figure*}

The power comparison of four methods under the two disease models is summerized in
Figures \ref{Fig:3} and \ref{Fig:4}. For the disease model where
the multiple phenotypes followed the same distribution, GSU had higher
power than the three SKAT-based methods (Figure \ref{Fig:3}). For
simulations with BBB phenotypes and GGG phenotypes, GSU attained higher
power than the other three methods when the sample size was 50, and
obtained substantial power improvement
as sample sizes increased. For the simulation with CCC phenotypes,
we also observed substantial power improvement of GSU than the other
three methods as sample size increased. Yet, the power of the three
SKAT-based methods were higher than that of GSU when the sample size was 50.
This can be explained by the inflated type 1 error rates of the SKAT-based
methods for the Cauchy-distributed phenotypes (Table \ref{Tab:2}).
Similarly, for disease models with a combination of different types
of phenotype distribution (i.e., BBG, BGG, and BGC), GSU had higher
power than SKAT, SKATO and AdjSKAT, regardless of sample size and
phenotype distributions (Figure \ref{Fig:4}). 

Comparing with univariate analysis in simulation I, GSU had further power improvement by considering the association of multi-phenotype in one single test. The three SKAT-based methods, on the other hand, performed multiple univariate analyses with Bonferroni correction. Under the alternative, the multiple phenotypes are correlated to each other. The univariate analysis with Bonferroni correction is subject to power loss without considering the correction structure of multiple phenotypes. Meanwhile, GSU can take the correlation structure into account when constructing the phenotype similarity, and therefore attained higher power than the other three methods.

\subsection{Computational Efficiency}

We compared the computational efficiency of R-based
GSU with SKAT, SKATO and AdjSKAT. We found GSU attained
highest computational efficiency among four methods. For example,
for the multivariate analysis of BBG phenotype with a sample size of 500 on
1000 simulation replicates, SKAT, SKATO and AdjSKAT took 5 (4320s),
16 (13686s) and 25 (20890s) times longer than GSU (849s),
using a personal computer with 2.3GHz CPU and 4G memory. The detailed
computational efficiency comparison of four methods for analyzing
data with various distributed phenotypes and different sample sizes
is given in Table S3 of Supplementary Materials. 

Although the comparisons were made under simulation setting, they represent general scenarios in practice. All four methods need to spend substantial time on eigen decomposition in the final step of calculating asymptotic distribution. For multivariate analysis, GSU performs the eigen decomposition once for all phenotypes, while SKAT-based methods need to perform the eigen decomposition for each phenotype. Moreover, if the link function is not identity link, iterative estimation for null model and additional large matrix multiplication and decomposition are required before the final eigen decomposition, which will also increase the computational burden of SKAT-based methods.

\section{Application to the Dallas Heart Study}

To evaluate the performance of GSU on real data, we applied it to
the Dallas Heart Study (DHS) sequence data\citep{Ahituv2007} and
compared the result of GSU with those of three SKAT-based methods.
The DHS sequencing data is comprised of 4 genes, \textit{ANGPTL3}, \textit{ANGPTL4},
\textit{ANGPTL5} and \textit{ANGPTL6}. After completing the quality control
(e.g., removing SNVs with high missing rate), 230 SNVs (54 SNVs, 63
SNVs, 61 SNVs, and 52 SNVs were from \textit{ANGPTL3}, \textit{ANGPTL4},
\textit{ANGPTL5} and \textit{ANGPTL6}, respectively) and 2598 subjects remained for the analysis.
In the real data analysis, we were interested in testing the association
of the SNVs in these genes with multiple metabolic-related phenotypes,
including obesity (dichotomized from BMI using a cut-off of 35), cholesterol,
high-density lipoprotein cholesterol (HDL), low-density lipoprotein
cholesterol (LDL) and very-low-density lipoprotein cholesterol (VLDL).

In order to consider the potential confounding effects of age, gender
and race, we adjusted these covariates in the analysis and used the
residuals to build phenotype similarity for GSU. Because the three
SKAT-based methods cannot directly analyze multiple phenotypes, we
obtained smallest p-values from the univariate analysis of all
phenotypes and then used the Bonferroni correction to determine whether
there was a significant association. The results are summarized in
Table \ref{Tab:3}. Because all 4 genes were metablolic candidate
genes, we first combined SNVs in the 4 genes into a single SNV-set.
For the joint analysis of 4 genes, GSU could detect a significant
association of 4 genes with 4 metabolic-related phenotypes (p-value=0.028),
while SKAT, SKATO and AdjSKAT did not detect the association. For
further exploration, we analyzed each gene separately and tested its
association with 4 metabolic-related phenotypes. By using GSU, we
found a marginal association of \textit{ANGPTL4} with the multiple phenotypes(p-value=0.057).

\begin{table*}[htbp]
\centering 

\caption{The multivariate analysis of 4 metabolic-related phenotypes in the
DHS study}

\begin{minipage}{130mm}

\label{Tab:3}

\begin{tabular}{cccccc}
\Hline
Gene  & Rare variants  & \multicolumn{4}{c}{P-value \footnote{Multiple phenotypes considered in the analysis include BMI,
cholesterol, HDL, LDL and VLDL. For SKAT-based methods, each p-value
listed in the table is the smallest p-value from 5 univariate analysis, where the Bonferroni corrected significance level should
be 0.05/5.}}\tabularnewline
\hline
 &  & GSU  & SKAT  & AdjSKAT  & SKATO\tabularnewline
\hline
\hline
ANGPTL3  & 54  & 0.300  & 0.083  & 0.071  & 0.112\tabularnewline
ANGPTL4  & 63  & 0.057  & 0.079  & 0.086  & 0.155\tabularnewline
ANGPTL5  & 61  & 0.075  & 0.046  & 0.073  & 0.107\tabularnewline
ANGPTL6  & 52  & 0.297  & 0.059  & 0.065  & 0.101\tabularnewline
\hline
ALL  & 230  & 0.028  & 0.250  & 0.174  & 0.245\tabularnewline
\hline
\end{tabular} 
\end{minipage}
\end{table*}

\section{Discussion}

Driven by recent developments in sequencing technologies and the common-diseases-rare-variants
hypothesis, sequencing studies are now emerging as a major study design
for the genetic association of complex diseases. Yet, the uniqueness
of sequencing data, including low MAF and high dimensionality, pose
daunting challenges to statistical analysis. The conventional methods,
such as a single-locus analysis using logistic regression, had low
power for sequencing data analysis. Instead, joint association analysis,
or SNV-set analysis, is becoming popular due to its ability to increase power and reduce multiple testing. Although
the existing joint association methods have nice features and are
easy to use, they also have certain limitations. For example, most
of burden tests assume homogeneous effects within the SNV-set, which
may not reflect the true underlying disease models. Besides burden
tests, most of the existing methods are parametric or semi-perimetric,
which often assume certain distributions of the phenotypes and mode
of inheritance. When the assumptions are violated, the results can
be unreliable.

To overcome these limitations, we have proposed a generalized similarity
U test for multivariate analysis of different types of phenotypes.
We conducted extensive simulation studies using data from the 1000
Genome Project and compared the performance of GSU with that of three other
popular methods: SKAT, AdjSKAT and SKATO. For all of the simulation
scenarios, including single phenotype with various distributions,
and multiple phenotypes with various combinations of phenotype distribution,
GSU outperformed the other three methods in terms of robust type I
error and higher power. Although the simulation results depend on
the simulation settings, and should always be interpreted in the context
of the simulation setting, we believe the results reflect the advantage
of GSU in a broader sense, because 1) the genetic data used in the
simulation comes from the 1000 Genome Project, which reflects the
LD pattern and the allelic frequency distribution in the general
population; and 2) we simulated a wide range of disease models to
mimic common disease scenarios.

In recent years, U-statistic-based methods have been popularly used
in genetic data analysis, and have shown their robustness and flexibility
for analyzing genetic data\citep{Schaid2005,Zhang2010,Wei2013}.
GSU is a general framework based on similarity measurements. Although
in this paper we used the weighted-IBS to calculate genetic similarity, other forms of genetic similarity can also be
used. For the weighted-IBS similarity, we can modify the weights (i.e., the original weight $w_{m}=1/\sqrt{\gamma_{m}(1-\gamma_{m})}$)
to reflect the importance of each SNV. Besides IBS-based similarity,
we can also use distance-transformed similarity to model the effect
in a nonlinear way. For example, we can use the Euclidean-Distance-based
similarity, $K_{i,j}^{ED}=exp(-\sum_{m=1}^{M}w_{m}(g_{i,m}-g_{j,m})^{2})$.
In this paper, we have focused on the analysis of categorical sequencing
data (i.e, SNV data). By using appropriate genetic similarity measurements,
GSU can easily be extended to analyze other types of genetic data,
such as count data (CNV data) and continuous data (expression data).

The flexibility of GSU also comes from the construction of the phenotype
similarity. By using phenotype similarity, GSU can analyze not only
a single phenotype, but also multiple phenotypes following different
distributions. Many genetic studies collect multiple secondary phenotypes,
or use intermediate biomarkers, to study complex diseases. By considering
multiple phenotypes that measure the different aspects of underlying
diseases, the power of association analysis can be potentially improved
\citep{Zhang2010,Liu2009,Maity2012}. Nevertheless, few methods have
been developed for multivariate analysis of sequencing data. Methods
were recently developed for multivariate analysis of common genetic variants,
but relies on the normal assumption for the phenotype distribution \citep{Maity2012}.
GSU is developed for both univariate and multivariate analysis of
sequencing data. It is distribution-free and can analyze multiple
phenotypes with a combination of different types of phenotype. Nevertheless, similar as all other non-parametric tests, it is not straightforward for GSU to perform covariate adjustment. One possible solution is to adopt the idea of covariate adjustment in regression, and include covariates when we calculate the centered similarity (e.g., $\tilde{S}=(I-X(X^TX)^{-1}X^T)S(I-X(X^TX)^{-1}X^T)$).

\backmatter





\section*{Supplementary Materials}

Supplementary Materials are available
with this paper at the Biometrics website on Wiley Online Library.
\vspace*{-8pt}

\bibliographystyle{biom} \bibliography{multi-trait}

\appendix
\label{appendix}


\section*{Appendix A: Centered Similarity}

\label{sub:Centered-Similarity}

By the following definition of centered similarity,
\begin{eqnarray*}
\tilde{h}(y_{1},y_{2})&=&h(y_{1},y_{2})-E(h(y_{1},Y_{2}))\\
& & -E(h(Y_{1},y_{2}))+E(h(Y_{1},Y_{2}))
\end{eqnarray*}
we can obtain conditional expectation for the centered phenotype similarity,
\begin{eqnarray*}
E(\tilde{h}(Y_{1},Y_{2})|Y_{1}) &=& E(h(Y_{1},Y_{2}|Y_{1}))-E(h(Y_{1},Y_{2}|Y_{1}))\\
& &-E(h(Y_{1},Y_{2}))+E(h(Y_{1},Y_{2}))\\
&=&0.
\end{eqnarray*}

Therefore, we have $E(\tilde{h}(Y_{1},Y_{2}))=0$ and $Var(E(\tilde{h}(Y_{1},Y_{2})|Y_{1}))=0$.
Using the same argument, we can have the same result for the centered
genetic similarity. 

Under the null hypothesis, when the genetic similarities are independent
of the phenotype similarities, we have,
\begin{eqnarray*}
E(U) & = & \frac{1}{n(n-1)}E(\sum_{i\neq j}\tilde{f}(G_{i},G_{j})\tilde{h}(Y_{i},Y_{j}))\\
 & = & \frac{1}{n(n-1)}\sum_{i\neq j}E(\tilde{f}(G_{i},G_{j}))E(\tilde{h}(Y_{i},Y_{j}))\\
 & = & 0.
\end{eqnarray*}


\section*{Appendix B: Proof of Theorem 1}
\label{sub:Proof-of-Thorem-1}

We can decompose the centered phenotype similarity by,
\[
\tilde{h}(y_{1},y_{2})=\sum_{s=1}^{\infty}\lambda_{s}\phi_{s}(y_{1})\phi_{s}(y_{2}),
\]
where $\{\lambda_{s}\}$ and $\{\phi_{s}(\cdotp)\}$ are eigenvalues and
eigenfunctions of the kennel $\tilde{h}(\cdotp,\cdotp)$.

Because of the orthogonality of $\{\phi_{s}(\cdotp)\}$ , we have
\begin{align*}
 & E(\tilde{h}(Y_{1},Y_{2})\phi_{s'}(Y_{2})|Y_{1})\\
= & \int\tilde{h}(Y_{1},y_{2})\phi_{s'}(y_{2})dF(y_{2})\\
= & \sum_{s=1}^{\infty}\lambda_{s}\phi_{s}(Y_{1})\int\phi_{s}(y_{2})\phi_{s'}(y_{2})dF(y_{2})\\
= & \lambda_{s'}\phi_{s'}(Y_{1}).
\end{align*}
We showed $E(\tilde{h}(Y_{1},Y_{2})\times1|Y_{1})=0\times1$
in Appendix A, which forced $\phi_{1}(\cdotp)=1$
and $\lambda_{1}=0$ to be an eigenfunction-eigenvalue pair in the
decomposition of $\tilde{h}(\cdotp,\cdotp)$. Again, because $\phi_{1}(\cdotp)$
is orthogonal with $\{\phi_{s}(\cdotp)\}_{s>1}$, for $s>1$, we have
\begin{eqnarray*}
E\phi_{s}(Y_{1}) & = & \int\phi_{s}(y_{1})\times1dF(y_{1})\\
 & = & \int\phi_{s}(y_{1})\phi_{1}(y_{1})dF(y_{1})\\
 & = & 0.
\end{eqnarray*}
Using the same argument, we have the corresponding results for
the decomposition of the centered genetic similarity:
\[
\tilde{f}(G_{i},G_{j})=\sum_{t=1}^{\infty}\eta_{t}\varphi_{t}(g_{1})\varphi_{t}(g_{2}).
\]
Then,
\begin{equation}
\begin{cases}
E\phi_{s}(Y)=0, & \forall\, s>1\\
E\varphi_{t}(G)=0 & \forall\, t>1.
\end{cases}\tag{Ax.1}\label{eq:E_eigen}
\end{equation}
Using the function decomposition, we can write GSU as,
\begin{eqnarray*}
U & = & \frac{1}{n(n-1)}\sum_{i\neq j}\tilde{f}(G_{i},G_{j})\tilde{h}(Y_{i},Y_{j})\\
 & = & \frac{1}{n(n-1)}\sum_{i\neq j}\sum_{t=1}^{\infty}\eta_{t}\varphi_{t}(G_{i})\varphi_{t}(G_{j})\sum_{s=1}^{\infty}\lambda_{s}\phi_{s}(Y_{i})\phi_{s}(Y_{j})\\
 & = & \frac{1}{n(n-1)}\sum_{t=1}^{\infty}\eta_{t}\sum_{s=1}^{\infty}\lambda_{s}\sum_{i\neq j}\varphi_{t}(G_{i})\varphi_{t}(G_{j})\phi_{s}(Y_{i})\phi_{s}(Y_{j})\\
 & = & \frac{1}{n-1}\sum_{t=2}^{\infty}\eta_{t}\sum_{s=2}^{\infty}\lambda_{s}\left(\frac{1}{\sqrt{n}}\sum_{i=1}^{n}\varphi_{t}(G_{i})\phi_{s}(Y_{i})\right)^{2}\\
 &  & -\frac{1}{n-1}\sum_{t=2}^{\infty}\eta_{t}\sum_{s=2}^{\infty}\lambda_{s}\frac{1}{n}\sum_{i=1}^{n}\left(\varphi_{t}(G_{i})\phi_{s}(Y_{i})\right)^{2}\\
 & = & \frac{1}{n-1}\sum_{t=2}^{\infty}\sum_{s=2}^{\infty}sign(\eta_{t}\lambda_{s})\left(\frac{1}{\sqrt{n}}\sum_{i=1}^{n}\eta_{t}^{\star}(G_{i})\phi_{s}^{\star}(Y_{i})\right)^{2}\\
 &  & -\frac{1}{n-1}\sum_{t=2}^{\infty}\sum_{s=2}^{\infty}sign(\eta_{t}\lambda_{s})\frac{1}{n}\sum_{i=1}^{n}\left(\eta_{t}^{\star}(G_{i})\phi_{s}^{\star}(Y_{i})\right)^{2},
\end{eqnarray*}
where $\varphi_{t}^{\star}(G_{i})=|\eta_{t}|^{0.5}\varphi_{t}(G_{i})$
and $\phi_{s}^{\star}(Y_{i})=|\lambda_{s}|^{0.5}\phi_{s}(Y_{i})$.

Under the null hypothesis, genotype ($G_{i}$) is independent of phenotypes
($Y_{i}$). Therefore, for $s>1$ and $t>1$,
\begin{align}
 & E(\eta_{t}^{\star}(G_{1})\phi_{s}^{\star}(Y_{1}))\nonumber \\
= & |\eta_{t}|^{0.5}E\varphi_{t}(G_{1})|\lambda_{s}|^{0.5}E\phi_{s}(Y_{1})\nonumber \\
= & 0,\tag{Ax.2}\label{eq:E2_eigen}
\end{align}
and
\begin{align}
 & E(\eta_{t}^{\star}(G_{1})\phi_{s}^{\star}(Y_{1})\eta_{t'}^{\star}(G_{1})\phi_{s'}^{\star}(Y_{1}))\nonumber \\
= & |\eta_{t}\lambda_{s}\eta_{t'}\lambda_{s'}|^{0.5}E(\varphi_{t}(G_{1})\varphi_{t'}(G_{1}))E(\phi_{s}(Y_{1})\phi_{s'}(Y_{1}))\nonumber \\
= & \begin{cases}
|\eta_{t}\lambda_{s}|, & \text{if \ensuremath{s=s'}and \ensuremath{t=t'}}\\
0, & \text{otherwise.}
\end{cases}\tag{Ax.3}\label{eq:E3_eigen}
\end{align}
Therefore, for any finite subset $\Delta$ of $\{(s,t)\}_{s>1,t>1}$,
$\left\{ \frac{1}{\sqrt{n}}\sum_{i=1}^{n}\eta_{t}^{\star}(G_{i})\phi_{s}^{\star}(Y_{i})\right\} _{(s,t)\in\Delta}$
converges to a multivariate normal distribution by using results from equation
\ref{eq:E2_eigen}, equation \ref{eq:E3_eigen} and multivariate CLT.

Additionally, we can show that,
\begin{align*}
 & \sum_{s>1,t>1}E(\eta_{t}^{\star}(G_{1})\phi_{s}^{\star}(Y_{1}))^{2}\\
= & \sum_{s}|\lambda_{s}|\sum_{t}|\eta_{t}|\\
< & \infty.
\end{align*}
Under the condition $\sum_{s>1,t>1}E(\eta_{t}^{\star}(G_{1})\phi_{s}^{\star}(Y_{1}))^{2}<\infty$,
the countable sequence of function $\{\eta_{t}^{\star}(\cdotp)\phi_{s}^{\star}(\cdotp)\}$
is a Donsker class\citep{Vaart2000}. Therefore, the empirical process,
$\frac{1}{\sqrt{n}}\sum_{i=1}^{n}\eta_{t}^{\star}(G_{i})\phi_{s}^{\star}(Y_{i})$,
converges weakly to the Gaussian process $Z_{s,t}$ with mean zero
and covariance function,
\begin{eqnarray*}
cov(Z_{s,t},Z_{s',t'})&=&E(\eta_{t}^{\star}(G_{1})\phi_{s}^{\star}(Y_{1})\eta_{t'}^{\star}(G_{1})\phi_{s'}^{\star}(Y_{1}))\\
&=&\begin{cases}
|\eta_{t}\lambda_{s}|, & \text{if \ensuremath{s=s'}and \ensuremath{t=t'}}\\
0, & \text{otherwise.}
\end{cases}
\end{eqnarray*}
With this uniform convergence (for all $s>1$ and $t>1$), we can
show that,
\begin{eqnarray*}
nU & \xrightarrow{D} & \sum_{t=2}^{\infty}\sum_{s=2}^{\infty}sign(\eta_{t}\lambda_{s})(Z_{s,t})^{2}\\
 &  & -\sum_{t=2}^{\infty}\sum_{s=2}^{\infty}sign(\eta_{t}\lambda_{s})|\eta_{t}\lambda_{s}|\\
 & = & \sum_{t=1}^{\infty}\eta_{t}\sum_{s=1}^{\infty}\lambda_{s}(\chi_{st}^{2}-1),
\end{eqnarray*}
where $\chi_{st}^{2}$ are i.i.d chi-squared random variables with
a d.f. of 1.

\section*{Appendix C: Proof of Theorm 2}
\label{sub:Proof-of-Theorm-2}

To simplify the notation, we denote $X=(Y,G)$ and $u(X_{1},X_{2})=\tilde{f}(G_{1},G_{2})\tilde{h}(Y_{1},Y_{2})$.
GSU can then be rewritten as:
\[
U=\frac{1}{n(n-1)}\sum_{i\neq j}u(X_{i},X_{j}).
\]
Define a centered kernel $\tilde{u}(x_{1},x_{2})$ by:
\[
\tilde{u}(x_{1},x_{2})=u(x_{1},x_{2})-u_{1}(x_{1})-u_{1}(x_{2})-\mu,
\]
where $u_{1}(x)=E(u(X_{1},X_{2})|X_{1}=x)$.

We can decompose the GSU as follows:
\begin{eqnarray*}
U & = & \frac{1}{n(n-1)}\sum_{i\neq j}u(X_{i},X_{j})\\
 & = & \frac{1}{n(n-1)}\sum_{i\neq j}\left(\tilde{u}(X_{i},X_{j})+u_{1}(X_{i})+u_{1}(X_{j})-\mu\right)\\
 & = & \frac{1}{n(n-1)}\sum_{i\neq j}\tilde{u}(X_{i},X_{j})+\frac{2}{n}\sum_{i=1}^{n}\left(u_{1}(X_{i})-\mu\right)+\mu.
\end{eqnarray*}
Thus,
\[
\sqrt{n}(U-\mu)=\frac{2}{\sqrt{n}}\sum_{i=1}^{n}\left(u_{1}(X_{i})-\mu\right)+\frac{\sqrt{n}}{n(n-1)}\sum_{i\neq j}\tilde{u}(X_{i},X_{j}).
\]
Becuase $E(u_{1}(X))=\mu$ and $Var(u_{1}(X))=\zeta_{1}$, the first
term converges to normal distribution by applying CLT:
\[
\frac{2}{\sqrt{n}}\sum_{i=1}^{n}\left(u_{1}(X_{i})-\mu\right)\xrightarrow{D}N(0,4\zeta_{1}).
\]
Then we need to show:
\[
R=\frac{\sqrt{n}}{n(n-1)}\sum_{i\neq j}\tilde{u}(X_{i},X_{j})\xrightarrow{p}0.
\]
This can be done by proving $ER^{2}\rightarrow0$, using the fact
that $E(\tilde{u}(X_{1},X_{2}))=0$, $Var(\tilde{u}(X_{1},X_{2}))<\infty$
and $E(\tilde{u}(X_{1},X_{2})|X_{1})=0$. In fact, by using the similar
technique in Appendix C, we can show that
$\sqrt{n}R$ asymptotically follows the distribution of a weighted
sum of independent chi-square random variables.

\section*{Appendix D: Matrix Similarity}
\label{sub:Matrix-Similarity}

In the study sample, we can calculate the centered phenotype similarity
by:
\[
\tilde{h}(y_{i},y_{j})=h(y_{i},y_{j})-\frac{1}{n}\sum_{j=1}^{n}h(y_{i},y_{j})-\frac{1}{n}\sum_{i=1}^{n}h(y_{i},y_{j})+\frac{1}{n^{2}}\sum_{i,j}h(y_{i},y_{j}).
\]
Denote $\tilde{S}_{i,j}=\tilde{h}(y_{i},y_{j})$ and $S_{i,j}=h(y_{i},y_{j})$, the above equation becomes:
\[
\tilde{S}_{i,j}=S_{i,j}-\frac{1}{n}\sum_{j=1}^{n}S_{i,j}-\frac{1}{n}\sum_{i=1}^{n}S_{i,j}+\frac{1}{n^{2}}\sum_{i,j}S_{i,j}.
\]
The equations can be written in a matrix form,
\begin{eqnarray*}
\tilde{S} & = & S-JS-SJ+JSJ\\
 & = & (I-J)S(I-J),
\end{eqnarray*}
where $\tilde{S}=\{\tilde{h}(y_{i},y_{j})\}_{n\times n}$, $S=\{h(y_{i},y_{j})\}_{n\times n}$,
$I=\{1_{\{i=j\}}\}_{n\times n}$, and $J=\{\frac{1}{n}\}_{n\times n}$.

Similarly, the centered genetic similarity can also be written in
the matrix form:
\[
\tilde{K}=(I-J)K(I-J),
\]

where $\tilde{K}=\{\tilde{f}(g_{i},g_{j})\}_{n\times n}$, and $K=\{f(g_{i},g_{j})\}_{n\times n}$.

\section*{Appendix E: Limiting Distribution}
\label{sub:Limiting-Distribution}

In the actual computation, we will use a matrix eigen-decomposition
to obtain the eigenvectors as a finite-dimension approximation
of the eigenfunctions. For a matrix eigen-decomposition, a computer
algorithm usually gives the eigenvalue $\lambda_{s}$ with the eigenvector
$\phi_{s}$, which satisfies $\sum_{i=1}^{n}\phi_{s,i}^{2}=1$ instead
of $\frac{1}{n}\sum_{i=1}^{n}\phi_{s,i}^{2}=1$. Therefore, using
the eigenvalues $\tilde{\lambda}_{s}$ and $\tilde{\eta}_{t}$ calculated
from the matrix eigen-decomposition, the limiting distribution of
GSU is:
\[
nU\sim\sum_{t=1}^{n}\frac{\tilde{\eta}_{t}}{n}\sum_{s=1}^{n}\frac{\tilde{\lambda}_{s}}{n}(\chi_{st}^{2}-1).
\]
.

\label{lastpage}

\end{document}